\documentclass[]{pasj01}

\Received{}
\Accepted{}
 
\usepackage{latexsym}
\usepackage[T1]{fontenc} 
\usepackage[switch,mathlines]{lineno}

\begin{document} 

\title{ Broadband non-thermal emission of odd radio circles induced by explosive galactic outflow remnants and their evolution
 }

\author{Yutaka \textsc{Fujita}\altaffilmark{1}%
}
\altaffiltext{1}{Department of Physics, Graduate School of Science,
Tokyo Metropolitan University\\
1-1 Minami-Osawa, Hachioji-shi, Tokyo 192-0397}
\email{y-fujita@tmu.ac.jp}

\author{Norita \textsc{Kawanaka},\altaffilmark{1,2}}
\altaffiltext{2}{National Astronomical Observatory of
Japan, 2-21-1 Osawa, Mitaka, Tokyo 181-8588}

\author{Susumu \textsc{Inoue}\altaffilmark{3,1}}
\altaffiltext{3}{International Center for Hadron Astrophysics, Chiba University\\
1-33 Yayoi-cho, Inage-ku, Chiba City, Chiba 263-8522}


\maketitle

\begin{abstract}
Odd radio circles (ORCs) are mysterious rings of faint, diffuse emission recently discovered in radio surveys, some of which may be associated with galaxies in relatively dense environments.
We propose such ORCs to be synchrotron emission from remnants of explosive galactic outflows, calling them OGREs, and discuss their broadband non-thermal emission and evolution.
We posit that a large amount of energy was ejected from the central galaxy in the past, creating an outgoing shock that accelerates cosmic rays.
Assuming plausible values for the density, temperature and magnetic field of the ambient medium, 
consistency with the observed spectral index, size and power of the ORCs requires the energy to be as high as $\sim 10^{60}$~erg,
suggesting that their sources could be active galactic nuclei.
We calculate the spectral energy distributions (SEDs) of the OGREs and their evolution, including synchrotron, inverse Compton (IC) and bremsstrahlung emission from electrons, and pion-decay emission from protons. 
We find that the SEDs of the younger OGREs are not greatly different from those of older ones currently observable as ORCs if radiative cooling of electrons is effective.
As such younger OGREs are expected to be rarer and smaller, they may not be readily observable.
However, if radiative cooling of electrons is ineffective, younger OGREs may be detectable in X-rays.
\end{abstract}


\section{Introduction} 
\label{sec:intro}

Odd radio circles (ORCs) were recently discovered in sensitive surveys by the Australian SKA Pathfinder Telescope (ASKAP), MeerKAT and Giant Metrewave Radio Telescope (GMRT) as diffuse radio emission with annular morphology and diameter of about one arcmin. Some of them may be associated with galaxies at their centers with redshifts $z \simeq 0.2$--$0.6$ \citep{2021Galax...9...83N,2021PASA...38....3N,2021MNRAS.505L..11K,2022MNRAS.513.1300N,2022MNRAS.512..265F,2022RNAAS...6..100O,2023arXiv230411784K,2023arXiv231206387R,2023arXiv231213298K,2024arXiv240101278K}.

Numerous proposals have been put forth on their origin and formation, such as supernova remnants in the Local Group \citep{2022MNRAS.512..265F,2022MNRAS.513L.101O,2023MNRAS.526.6214S}, galactic wind termination shocks \citep{2022MNRAS.513.1300N}, black hole mergers \citep{2021MNRAS.505L..11K}, tidal disruptions of stars by a supermassive black hole \citep{2022MNRAS.516L..43O}, galaxy mergers \citep{2023ApJ...945...74D}, virial shocks around massive galaxies \citep{2023arXiv230917451Y}, jet-inflated bubbles \citep{2024arXiv240108207L}, and radio galaxies \citep{2024arXiv240209708S}.
In particular, ORC4 has a relatively massive elliptical galaxy in its center that has recently been found to be surrounded by an atypically large amount of ionized gas. This, together with ORC4 itself, has been interpreted as resulting from the past activity of an energetic galactic wind \citep{2023arXiv231015162C}.

The first goal of this paper is to show that explosive injection of energy at the centers of galaxies can provide a plausible explanation for the ORCs associated with galaxies, in particular ORC~4 and ORC~1. We suggest that they are remnants of galactic outflows (OGREs)\footnote{OGREs = Outflow-from-Galaxy REmnants}, in which a shock is generated by an energetic, short-lived event in the central galaxy and propagates out to large radii corresponding to such ORCs. We study the expected broadband non-thermal emission due to cosmic rays (CRs) accelerated by the shock. 
The second goal is to make predictions for future observations in other wavebands for the known ORCs, as well as OGREs in their younger stages, considering the time evolution of the shock and the non-thermal emission.
We assume $H_0=70\rm\: km\: s^{-1}\: Mpc^{-1}$, $\Omega_m=0.3$, and $\Omega_\Lambda=0.7$.

\section{Models}
\label{sec:model}

Concerning the dynamical evolution of the shock and CR acceleration,
we adopt the formulation of \citet{2007ApJ...663L..61F}, which is based on a model for supernova remnants \citep{2006MNRAS.371.1975Y}. 
Our main focus is on ORC4 \citep{2021PASA...38....3N,2023arXiv231015162C} and ORC1 \citep{2022MNRAS.513.1300N} that have elliptical galaxies at their centers and have been observed in multiple radio bands.
The redshift and stellar mass of ORC4's central galaxy are $z = 0.4512$ and $M_*\sim 1.9\times 10^{11}\: M_\odot$, respectively
\citep{2023arXiv231015162C}.
Those of ORC1 are $z=0.551$ and $M_* \sim 2.8\times 10^{11}\: M_\odot$, respectively \citep{2022MNRAS.513.1300N}.

We assume that the galaxy and its environment is spherically symmetric and that the density profile of the circumgalactic medium (CGM) has a power-law form:
\begin{equation}
\label{eq:ISM}
 \rho(r) = \rho_1(r/r_1)^{-\omega}\:,
\end{equation}
where $r$ is the distance from the galactic center.
We set $\rho_1=7.0\times 10^{-27}\rm\: g\: cm^{-3}$, $r_1=10$~kpc, and $\omega=1.43$ based on the observations of the nearby elliptical galaxy NGC~4636 \citep{2021MNRAS.506.2030M}, whose stellar mass estimated from the $V$-band luminosity ($M_*=4.72\times 10^{11}\: M_\odot$) is comparable to the central galaxies of ORC4 and ORC1.
Although the density at $r \sim $200 kpc relevant to ORCs is not directly constrained from observations, we assume that this profile extends to such scales, which may be consistent with the observed fact that some ORCs lie in overdense environments \citep{2021Galax...9...83N} where the presence of an associated intragroup medium is likely.

For simplicity, we assume that energy was ejected from the galactic center on a time scale much shorter than the current age of the shock.
Using a shell approximation (e.g. Section~III in \cite{1988RvMP...60....1O}), the radius of the shock can be written as
\begin{equation}
\label{eq:Rs}
 R_s = \xi\left(\frac{{\cal E}_0}{\rho_1 r_1^{\omega}}\right)
^{1/(5-\omega)}t^{2/(5-\omega)}\:,
\end{equation}
where ${\cal E}_0$ is the ejected energy, and $t$ is the time since the explosion. 
The indices can be determined so that the coefficient $\xi\: (\sim 1)$ is non-dimensional. 
Equation~(\ref{eq:Rs}) can be used when the ambient temperature (pressure) is small enough that the shock Mach number $\mathcal{M}$ is greater than one. In fact, regardless of the temperature of the ambient medium, numerical simulations indicate that equation~(\ref{eq:Rs}) describes the evolution of the shock well as long as $\mathcal{M}>1$ (e.g. \cite{1990ApJ...354..468M}).

The emission from ORCs is most likely synchrotron radiation by relativistic electrons \citep{2021Galax...9...83N,2021PASA...38....3N}.
If the electrons are accelerated at shocks via diffusive shock acceleration (DSA), the spectral index $\alpha$ of the synchrotron emission from freshly accelerated electrons can be related to the shock Mach number as
\begin{equation}
\label{eq:alpha}
 \alpha = \frac{\mathcal{M}^2 + 1}{\mathcal{M}^2 - 1} - \frac{1}{2}\:,
\end{equation}
as long as the simple, test particle approximation for DSA is valid
\citep{1978MNRAS.182..147B,1983RPPh...46..973D,1987PhR...154....1B,2013PASJ...65...16A}.
The test particle approximation may be reasonable for shocks with sufficiently low Mach numbers, such as in merging galaxy clusters
where the relation above is observationally supported
\citep{2021MNRAS.506..396W}.
We will show later that the shocks corresponding to ORCs in the OGRE scenario also has similarly low Mach numbers.
Thus, from the observationally inferred value of $\alpha$, 
we can estimate the shock velocity $V_s$ if the sound velocity $c_s$ of the ambient medium is assumed.

The observed radio spectral index $\alpha_{\rm obs}$ can differ from $\alpha$ in equation~(\ref{eq:alpha}) if the radiating electrons are significantly affected by synchrotron and inverse Compton cooling. Since the cooling time is generally expected to be shorter than the age of the ORCs (see below), fiducially we assume that the observed emission results from cooled electrons downstream of the shock, so that $\alpha_{\rm obs}=\alpha+0.5$ (e.g. \cite{1999ApJ...520..529S}.)

Since $V_s=\dot{R_s}$, $t$ is represented as
\begin{equation}
\label{eq:time}
 t = \frac{2}{5-\omega}\frac{R_s}{V_s}\:.
\end{equation}
from equation~(\ref{eq:Rs}). Substituting this into equation~(\ref{eq:Rs}),
the energy is derived as
\begin{equation}
\label{eq:E0}
 {\cal E}_0 = \rho_1 r_1^{\omega}\xi^{-(5-\omega)}\left[\frac{(5-\omega)\mathcal{M} c_s}{2}\right]^2 R_s^{3-\omega}\:.
\end{equation}

We assume that the energy distributions of the freshly accelerated CR particles are described by
\begin{equation}
\label{eq:Ni(E)}
 N_i(E) \propto E^{-x}e^{-E/E_{{\rm max},i}}\:,
\end{equation}
where $E$ is the particle energy, and $E_{{\rm max},i}$ is the maximum energy for protons ($i=p$) and electrons ($i=e$). 
The index is represented by $x = (r_b + 2)/(r_b - 1)$, where 
\begin{equation}
\label{eq:rb}
 r_b = \frac{(\gamma+1)\mathcal{M}^2}{(\gamma-1)\mathcal{M}^2 + 2}
\end{equation}
is the compression ratio of the shock (e.g. \cite{1987PhR...154....1B}).

The maximum energies for CR protons and electrons are given by solving the following equations
\begin{equation}
\label{eq:tacc}
 t_{\rm acc} = \min(t_{pp},t),~~~t_{\rm acc} = \min(t_{\rm synIC},t)\:,
\end{equation}
respectively. Here $t_{\rm acc}$, $t_{pp}$, $t_{\rm synIC}$ are the acceleration time, the cooling time of protons due to interactions with  ambient protons ($pp$ interaction), and the cooling time of electrons by synchrotron and inverse Compton (IC) losses, respectively.

For DSA, the acceleration time is
given as
\begin{equation}
 t_{\rm acc} = \frac{20 \eta cE_{{\rm max},i}}{e B_d V_s^2}\:,
\end{equation}
where $c$ is the speed of light, and $e$ is the elementary charge \citep{1987ApJ...313..842J,2004A&A...416..595Y}. We assume $\eta=1$ according to \citet{2007ApJ...663L..61F}. The magnetic field downstream of the shock is
$B_d=r_b B$, where $B$ is the upstream magnetic field strength, assumed to be given by
\begin{equation}
\label{eq:B(r)}
 B(r)=B_2(\rho(r)/\rho(r_2))^{2/3}\:,
\end{equation}
where $r_2=50$~kpc. This relation between $B$ and $\rho$ is valid when the magnetic fields are relatively ordered and frozen in to the plasma, so that $B^2 \propto L^{-2}$ while $\rho\propto L^{-3}$, where $L$ is the scale length of a fluid element \citep{2018SSRv..214..122D}. We assume this to be the case for the CGM at $r_2 > 50$~kpc.
The proton lifetime is
\begin{equation}
 t_{pp} = 5.3\times 10^7\: (n_{\rm CGM}/{\rm cm^{-3}})^{-1}{\:\rm yr}\:,
\end{equation}
where $n_{\rm CGM}$ is the CGM number density. In our calculations, proton cooling due to $pp$ interactions is negligible, as $t<10^9$~Gyr (see Section~\ref{sec:result})
and $n_{\rm CGM}<0.01\rm\: cm^{-2}$ implies that $t_{pp} > t$.

The power of synchrotron and IC emission from an electron of energy $E$ are given respectively by $P_{\rm syn}\propto E^2 U_B$ and $P_{\rm IC}\propto E^2 U_{\rm ph}$ in the Thomson regime, where $U_B$ is the magnetic energy density and $U_{\rm ph}$ is the photon energy density \citep{1979rpa..book.....R}.
Since the synchrotron-IC cooling time $t_{\rm synIC}$ satisfies $E/t_{\rm synIC}\propto P_{\rm syn} + P_{\rm IC}$,
\begin{eqnarray}
\label{eq:synIC}
 t_{\rm synIC} &=& 
1.25\times 10^6\: \left(\frac{E}{10\rm\: TeV}\right)^{-1}
\nonumber\\
& \times&\frac{1}{(B_d/{\rm\mu G})^2 + (B_{\rm CMB}/{\rm\mu G})^2}{\:\rm yr}\:,
\end{eqnarray}
where $B_{\rm CMB}=3.24\: (1+z)^2\rm\mu
G$ is the equivalent magnetic field strength for energy losses due to IC
scattering with cosmic microwave background (CMB) photons. 
The electron break energy where $t=t_{\rm synIC}$ is 
\begin{eqnarray}
 \label{eq:Ebr}
 E_{\rm br} &=& 
0.125\: \left(\frac{t}{10^8\rm\: yr}\right)^{-1}
\nonumber\\
& \times&\frac{1}{(B_d/{\rm\mu G})^2 + (B_{\rm CMB}/{\rm\mu G})^2}{\:\rm TeV}\:,
\end{eqnarray}
which is generally smaller than $E_{{\rm max},e}$ in our model.  For
$E > E_{\rm br}$ ($E < E_{\rm br}$), the cooling time $t_{\rm
synIC}$ is shorter (longer) than the shock age $t$. 
From relations~(\ref{eq:tacc}), we have
\begin{equation}
 \label{eq:Emaxp}
 E_{{\rm max},p}=1.6\times 10^4
\left(\frac{V_s}{10^3\rm\: km\: s^{-1}}\right)^2
\frac{B_d}{\rm \mu G}\frac{t}{10^8\rm\: yr}\:\rm TeV\:,
\end{equation}
\begin{eqnarray}
 \label{eq:Emaxe}
 E_{{\rm max},e}&=&44\:\frac{V_s}{10^3\rm\: km\: s^{-1}}\nonumber\\
&\times& \frac{(B_d/{\rm \mu G})^{1/2}}{((B_d/{\rm\mu G})^2 + (B_{\rm CMB}/{\rm\mu G})^2)^{1/2}}
\:\rm TeV\:.
\end{eqnarray}
The minimum energies of protons ($E_{{\rm min},p}$) and ($E_{{\rm min},e}$) electrons are assumed to be their rest mass energies.

The normalization of the particle distributions depends on the total energy of the CR protons within the shock $\epsilon {\cal E}_0$, where $\epsilon$ is a parameter, and the ratio of CR electrons to CR protons $K_{\rm ep}$ freshly accelerated at the shock (see Appendix).

Due to cooling, the energy distribution of radiating electrons steepens at $E\gtrsim E_{\rm br}$.
Thus, in steady state, we adopt for the electron distribution
\begin{eqnarray}
N_{e,c}(E)&=& \frac{A_e E_{\rm br}}{x-1}E^{-(x+1)}e^{-E/E_{{\rm max},e}} \nonumber\\
&&\times\left\{
\begin{array}{ll}
1-(1-E/E_{\rm br})^{x-1},& E< E_{\rm br}\\
1                       ,& E\geq E_{\rm br}
\end{array}
\right.
\end{eqnarray}
where $A_e$ is give in Appendix \citep{1999ApJ...520..529S}. Note that $N_{e,c}(E)$ approaches $A_e E^{-x}e^{-E/E_{{\rm max},e}}$ at $E\ll E_{\rm br}$.

Once the distributions are determined, we calculate the synchrotron, bremsstrahlung, and IC emissions of electrons, and the $\pi^0$ decay gamma rays from $pp$ interactions. 
Radiation processes for electrons are calculated using the models by \citet{2008MNRAS.384.1119F}, and gamma-ray spectra are derived based on the models by \citet{2006ApJ...647..692K}, \citet{2006PhRvD..74c4018K}, and \citet{2008ApJ...674..278K}.
The CR protons interact with the high density shell behind the shock with a mass of
\begin{equation}
  M_{\rm sh}=\int_0^{R_s}4\pi r^2 \rho(r) dr
\end{equation} 
The emission from secondary electrons/positrons produced in $pp$ interactions is negligible. 

\begin{figure*}[ht!]
 \begin{center}
  \includegraphics[width=8cm]{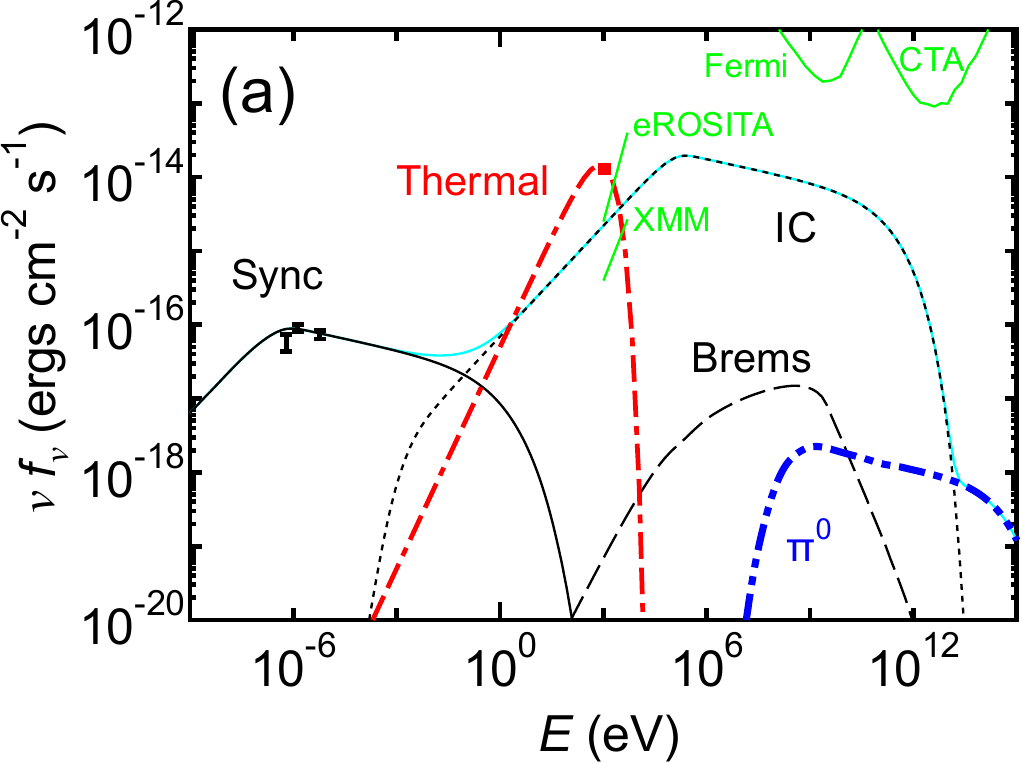} 
  \includegraphics[width=8cm]{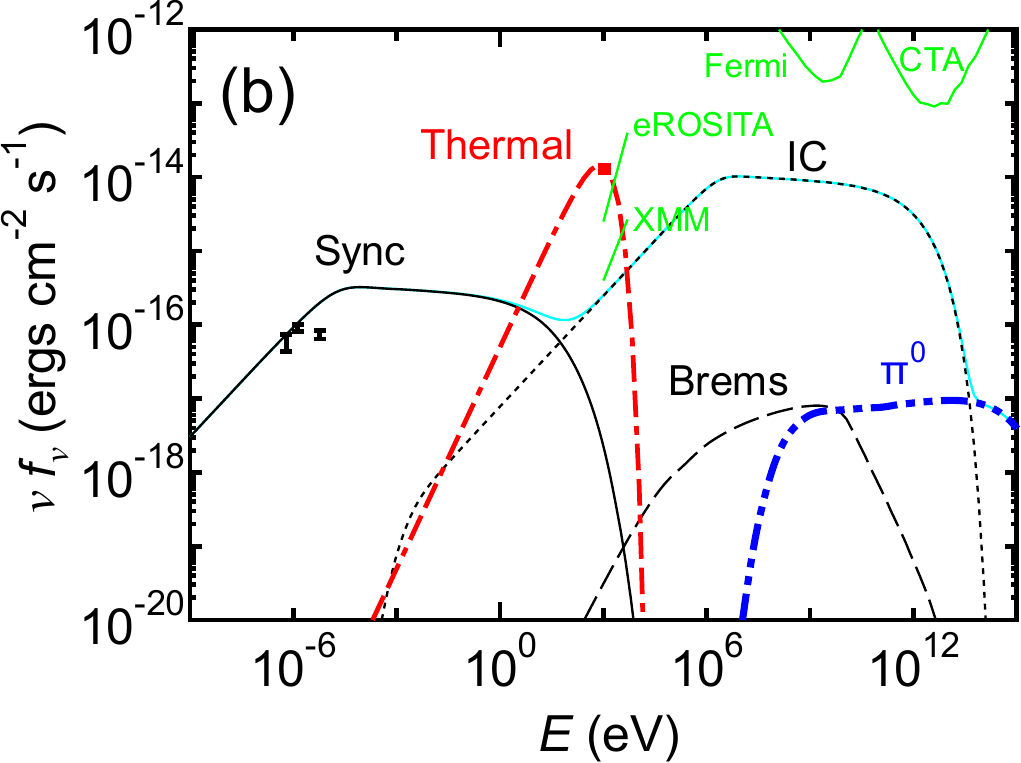} 
 \end{center}
\caption{SED of ORC4 compared to the OGRE model, the thin black solid, dashed and dotted lines represent the
synchrotron, non-thermal bremsstrahlung and IC scattering, respectively. 
The thick blue two-dotted–dashed line is for the $\pi^0$ decay gamma rays. 
Radio observations are represented by the black dots \citep{2021PASA...38....3N}. The sum of the four emissions is indicated by the cyan line. The expected thermal emission is indicated by the red dotted-two-dashed line (see text). The sensitivities of XMM-Newton, eROSITA, Fermi and CTA are indicated by the green lines. (a) Present time ($t_0=66$~Myr). (b) $t=0.2\: t_0=13$~Myr.
\label{fig:ORC4}}
\end{figure*}

In addition to $\epsilon$ and $K_{\rm ep}$, which will be determined later, we need to specify $\alpha$ (equation~(\ref{eq:alpha})), $R_s$ and $c_s$ (equation~(\ref{eq:E0})), and $B_2$ (equation~(\ref{eq:B(r)})). The parameters $\alpha$ and $R_s$ are obtained from observations. 
The sound velocity depends on the gas temperature $kT$:
\begin{equation}
 c_s = \sqrt{\gamma\frac{kT}{\mu m_P}}
=365\left(\frac{kT}{0.5\rm\: keV}\right)^{1/2}\rm\: km\: s^{-1}\:,
\end{equation}
where $m_p$ is the proton mass. For NGC~4636, the observed temperature is $kT\sim 0.8$~keV at $r\lesssim 20$~kpc. Since the temperature at $r\sim R_s\sim 200$~kpc is unknown, we assume $kT=0.5$~keV, which is comparable to that of the intragroup medium of galaxy groups, as may be expected around ORCs lying in overdense environments \citep{2021Galax...9...83N}. 
Although larger $B_2$ would allow larger synchrotron luminosity, we limit its value to $B_2\lesssim4.5\rm\: \mu G$ to avoid the unlikely situation of the magnetic energy exceeding the thermal energy of the gas at the shock.

\section{Model parameters}

In summary, we use three observables of ORCs as input parameters: (1) the size $R_s$, (2) the radio luminosity, and (3) the spectral index $\alpha$. In addition, we assume (4) the gas temperature $kT$, (5) the power law index characterizing the density profile $\omega$, (6) the magnetic field $B_2$, (7) the fraction of shock energy in CR protons $\epsilon$, and (8) the ratio of CR electrons to CR protons $K_{\rm ep}$.
The plausibility of the model will be demonstrated by (a) finding explosion energies ${\cal E}_0$ that are consistent with those of active galactic nuclei (AGNs), (b) allowing sufficient time for the relativistic electrons to be accelerated by DSA, and (c) allowing the relativistic electrons to have a long enough lifetime that the source is still visible.

Among the above three input parameters, the spectral index $\alpha$ is the most critical because it determines the overall shape of the non-thermal spectra. The parameters that are not directly constrained by observations mainly affect the normalization of the spectra. While they must be adjusted to reproduce the observations, given the currently limited information on ORCs, some of the parameters are likely to be degenerate with each other. Thus, we adopt values that we consider to be reasonable and exemplary. The dependence on the parameters could be summarized as follows. The proton energy fraction $\epsilon$ and the electron-proton ratio $K_{\rm ep}$ obviously affect the normalization. The magnetic field strength affects the synchrotron flux. Temperatures much smaller than $kT=0.5$~keV would lead to ${\cal E}_0$ (equation~(\ref{eq:E0})) that is insufficient to reproduce the observed radio luminosities of the ORCs. If we significantly increase the index $\omega$ in equation~(\ref{eq:ISM}) from 1.43, the non-thermal flux decreases, mainly because of the factor $(r_1/R_s)^\omega$ in equation~(\ref{eq:E0}). These indicate that the CGM temperature and density around the ORCs must be relatively high, which is consistent with the observed fact that some reside in overdense environments. 

\begin{table*}
  \tbl{Parameters.}{%
  \begin{tabular}{ccccccccccc}
      \hline
 & Cooling\footnotemark[$*$] & $R_s$ & $t$ & $\alpha$ & $\mathcal{M}$ & $V_s$ & $B_2$ & $\epsilon$ & $K_{\rm ep}$ & ${\cal E}_0$ \\
 &  & (kpc) & (Myr) &  &  & ($\rm km\: s^{=1}$) & ($\rm \mu G$) &  &  & ($10^{60}$~erg) \\
      \hline
ORC4 & Yes & 200 & 66 ($=t_0$) & 0.6 & 4.6 & 1670 & 0.5 & 0.1 & 0.01 & 2.9 \\
ORC4 & Yes & 81 & 13 & 0.52 & 9.3 & 3390 & 0.5 & 0.1 & 0.01 & 2.9 \\
ORC1 & Yes & 260 & 160 ($=t_0$) & 0.9 & 2.4 & 894 & 1.2 & 0.1 & 0.01 & 1.3 \\
ORC1 & Yes & 106 & 32 & 0.58 & 5.0 & 1810 & 1.2 & 0.1 & 0.01 & 1.3 \\
ORC4 & No & 200 & 130 ($=t_0$) & 0.92 & 2.4 & 876 & 4.5 & 0.1 & 0.05 & 0.81 \\
ORC4 & No & 81 & 25 & 0.59 & 4.9 & 1780 & 4.5 & 0.1 & 0.05 & 0.81 \\
ORC1 & No & 260 & 220 ($=t_0$) & 1.4 & 1.8 & 655 & 4.5 & 0.3 & 0.1 & 0.68 \\
ORC1 & No & 106 & 44 & 0.66 & 3.6 & 1330 & 4.5 & 0.3 & 0.1 & 0.68 \\
      \hline
    \end{tabular}}\label{tab:par}
\begin{tabnote}
\footnotemark[$*$] With or without radiative cooling.  \\ 
\end{tabnote}
\end{table*}

\begin{figure*}[ht!]
 \begin{center}
  \includegraphics[width=8cm]{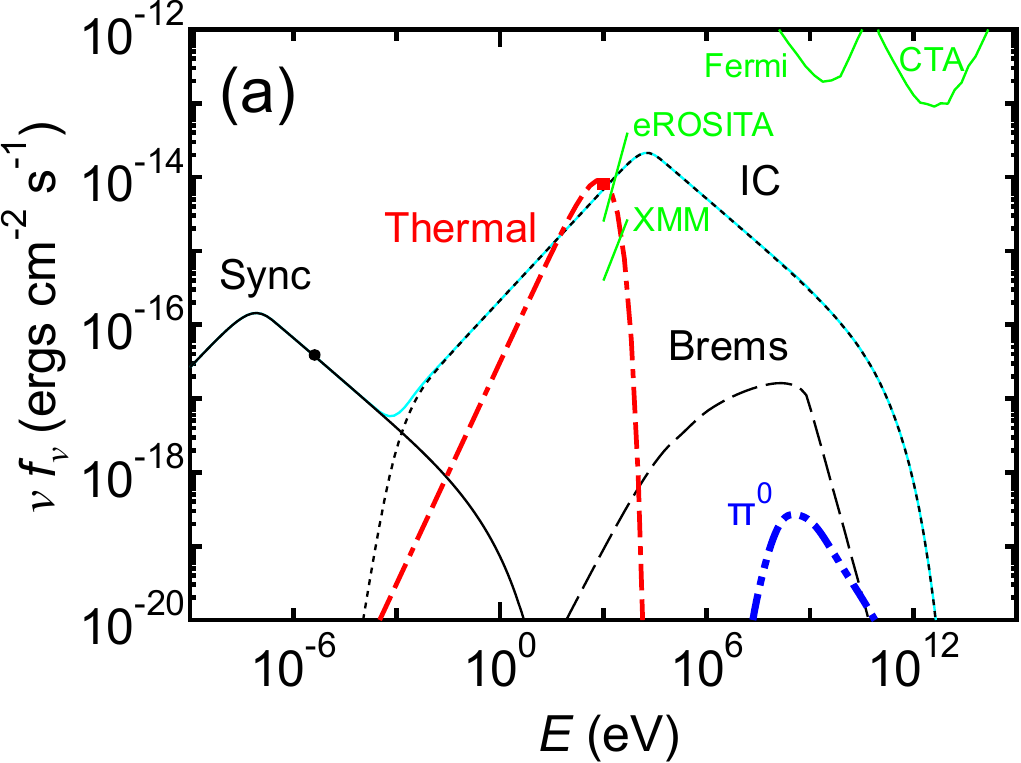} 
  \includegraphics[width=8cm]{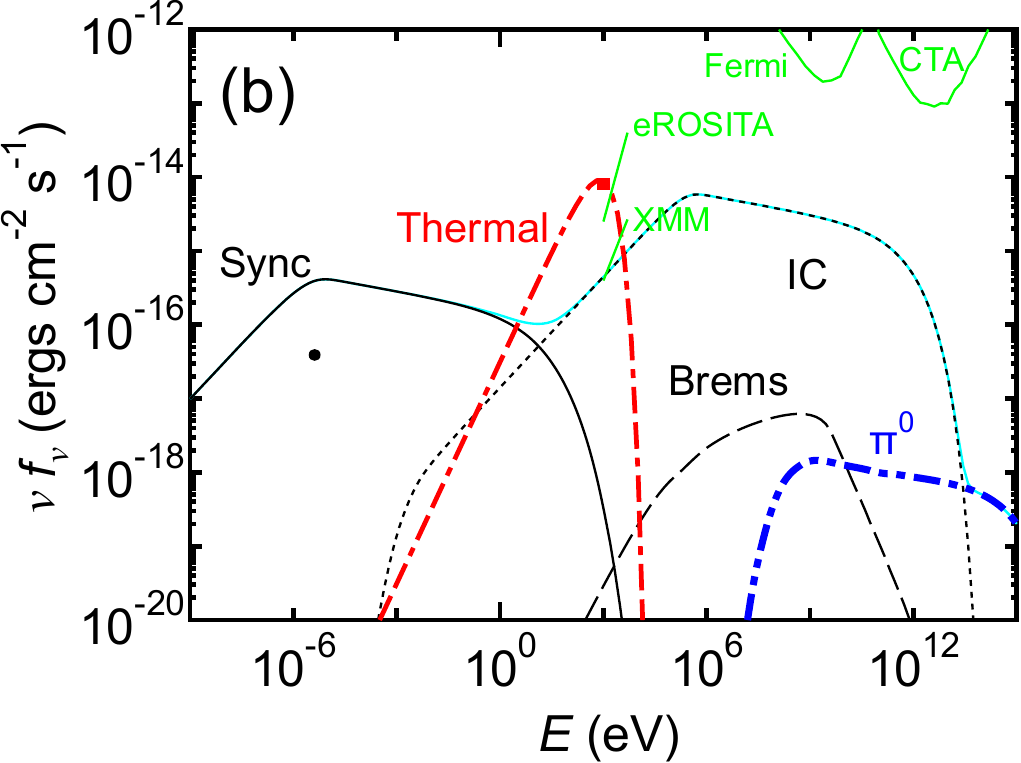} 
 \end{center}
\caption{Same as figure~\ref{fig:ORC4} but for ORC1. Radio observations are shown by the black dot \citep{2022MNRAS.513.1300N}. (a) Current time ($t_0=160$~Myr). (b) $t=0.2\: t_0=32$~Myr.\label{fig:ORC1}}
\end{figure*}

\section{Results}
\label{sec:result}

\subsection{ORC4}
\label{sec:ORC4}

The observed radio spectral index of ORC4 is $0.92\pm 0.18$ \citep{2021PASA...38....3N}. 
If we take $\alpha_{\rm obs}=0.92$, $\alpha=\alpha_{\rm obs}-0.5=0.42$ and $x = 2\alpha + 1 = 1.84$ \citep{1979rpa..book.....R}. This value of $x$ ($\alpha$) is less than 2 (0.5), which corresponds to the lower bound for DSA in the test particle approximation ($\mathcal{M}\rightarrow \infty$ in equation~(\ref{eq:alpha})). Thus, we choose to take $\alpha_{\rm obs}=1.1$, the maximum value allowed from observations, which implies $\alpha=0.6$, $\mathcal{M}=4.6$ (equation~(\ref{eq:alpha})), and $x=2.2$. These values of $\alpha_{\rm obs}$ and $\alpha$ adequately describe the observed radio spectrum (see below). Note that $\alpha_{\rm obs}$ ($\alpha$) should not be taken too close to 1 (0.5) in our formulation, since $\mathcal{M}$ and ${\cal E}_0$ would then diverge (equations~(\ref{eq:alpha}) and~(\ref{eq:E0})).

The shock radius is $R_s=200$~kpc \citep{2022MNRAS.513.1300N} and the velocity is $V_s=1670\rm\: km\: s^{-1}$. We assume $\epsilon=0.1$ and $K_{\rm ep}=0.01$ (table~\ref{tab:par}). Such values of $\epsilon$ is typically assumed for supernova remnants ($0.05\lesssim \epsilon \lesssim 0.5$; e.g. \cite{2013AandARv..21...70B}) and that of 
$K_{\rm ep}$ is comparable to that observed in Galactic CRs ($\sim 0.01$; \cite{2002cra..book.....S}).
We assume $B_2=0.5\:\rm \mu G$ so that the predicted radio flux matches the observed one.

Figure~\ref{fig:ORC4}(a)\footnote{References for detector sensitivities are https://xmm-tools.cosmos.esa.int/external/xmm\_user\_support/documentation/uhb/basics.html (XMM-Newton; 100~ks), \citet{2021AandA...647A...1P} (eROSITA; 30~ks), https://www.slac.stanford.edu/exp/glast/groups/canda/lat\_Performance.htm (Fermi; 10~yr), and https://www.cta-observatory.org/science/ctao-performance/ (CTA; 50~h). The curves for XMM-Newton and eROSITA are for point sources, so the actual sensitivities for OGREs with source extensions could be worse.} shows the current SED for ORC4. The synchrotron (IC) emission dominates at lower (higher) energies. The explosion energy is ${\cal E}_0=2.9\times 10^{60}$~erg from equation~(\ref{eq:E0}). The time since the explosion is estimated to be $t_0=66$~Myr from equation~(\ref{eq:Rs}). The electron break energy is $E_{\rm br}=4.1$~GeV from equation~(\ref{eq:Ebr}). The maximum proton and electron energies are $E_{{\rm max},p}=14$~PeV and $E_{{\rm max},e}=7.4$~TeV, respectively (equations~(\ref{eq:Emaxp}) and~(\ref{eq:Emaxe})).
The cooling break in the synchrotron emission turns out to be close to the observed frequency bands ($\sim 10^{-5}$~eV) . Thus the electrons emitting at these frequencies may not yet be fully cooled ($E\sim E_{\rm br}$), so that $\alpha_{\rm obs}-0.5<\alpha<\alpha_{\rm obs}$. Thus, even if $\alpha_{\rm obs}=0.92$ with a single power-law fit to the observations, $\alpha>0.5$ is allowed.

Figure~\ref{fig:ORC4}(b) shows the SED at $t=0.2\: t_0=13$~Myr. The radius is $R_s=81$~kpc (equation~(\ref{eq:Rs})) and the velocity is $V_s=3390\rm\: km\: s^{-1}$ ($\mathcal{M}=9.3$). The electron break energy is $E_{\rm br}=20$~GeV. The maximum energies are $E_{{\rm max},p}=29$~PeV and $E_{{\rm max},e}=24$~TeV.
The broadband SED is not greatly different compared to that at $t=t_0$ in figure~\ref{fig:ORC4}(a), because the values of $\alpha$ are similar ($\alpha=0.52$ compared to $\alpha=0.6$).

\subsection{ORC1}

The observed radio spectral index of ORC1 is $1.4\pm 0.05$ \citep{2022MNRAS.513.1300N} and we assume $\alpha_{\rm obs}=1.4$, which means $\alpha=0.9$, $\mathcal{M}=2.4$ (equation~(\ref{eq:alpha})) and $x=2.8$. The shock radius is $R_s=260$~kpc \citep{2022MNRAS.513.1300N} and the velocity is $V_s=894\rm\: km\: s^{-1}$. We set $\epsilon=0.1$ and $K_{\rm ep}=0.01$ as in ORC4 (table~\ref{tab:par}). We assume that $B_2=1.2\:\rm \mu G$ so that the predicted radio flux matches that observed. The explosion energy is ${\cal E}_0=1.3\times 10^{60}$~erg (equation~(\ref{eq:E0})) and the time since the energy ejection is estimated to be $t_0=160$~Myr (equation~(\ref{eq:Rs})). The electron break energy is $E_{\rm br}=1.3$~GeV. 
The maximum energies are $E_{{\rm max},p}=14$~PeV and $E_{{\rm max},e}=4.1$~TeV  (equations~(\ref{eq:Emaxp}) and~(\ref{eq:Emaxe})).

Figure~\ref{fig:ORC1}(a) shows the current SED. 
Here the observed radio emission is at frequencies appreciably above the cooling break of the synchrotron emission.
Figure~\ref{fig:ORC1}(b) shows the SED at $t=0.2\: t_0=32$~Myr. The radius is $R_s=106$~kpc (equation~(\ref{eq:Rs})) and the velocity is $V_s=1810\rm\: km\: s^{-1}$. The electron break energy is $E_{\rm br}=6.0$~GeV. The maximum energies are $E_{{\rm max},p}=35$~PeV and $E_{{\rm max},e}=14$~TeV. 
The SED is similar to those of ORC4 (figure~\ref{fig:ORC4}), due to the larger Mach number ($\mathcal{M}=5.0$) and the harder spectra ($\alpha=0.58$ and $x=2.2$),

\begin{figure*}[ht!]
 \begin{center}
  \includegraphics[width=8cm]{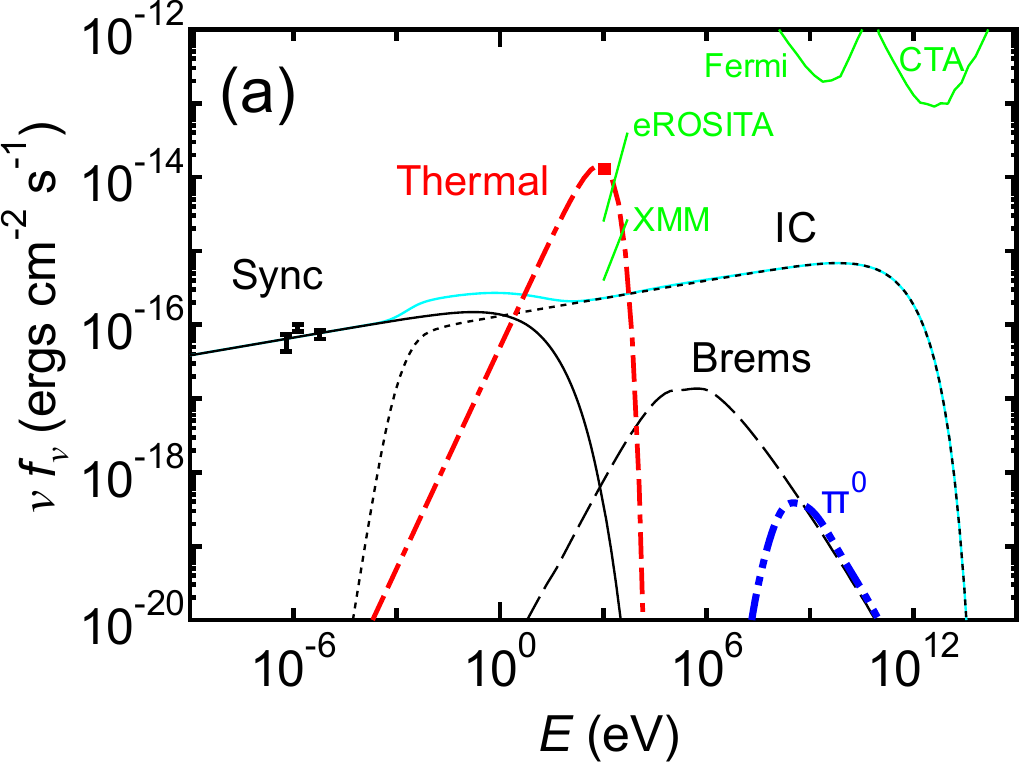} 
  \includegraphics[width=8cm]{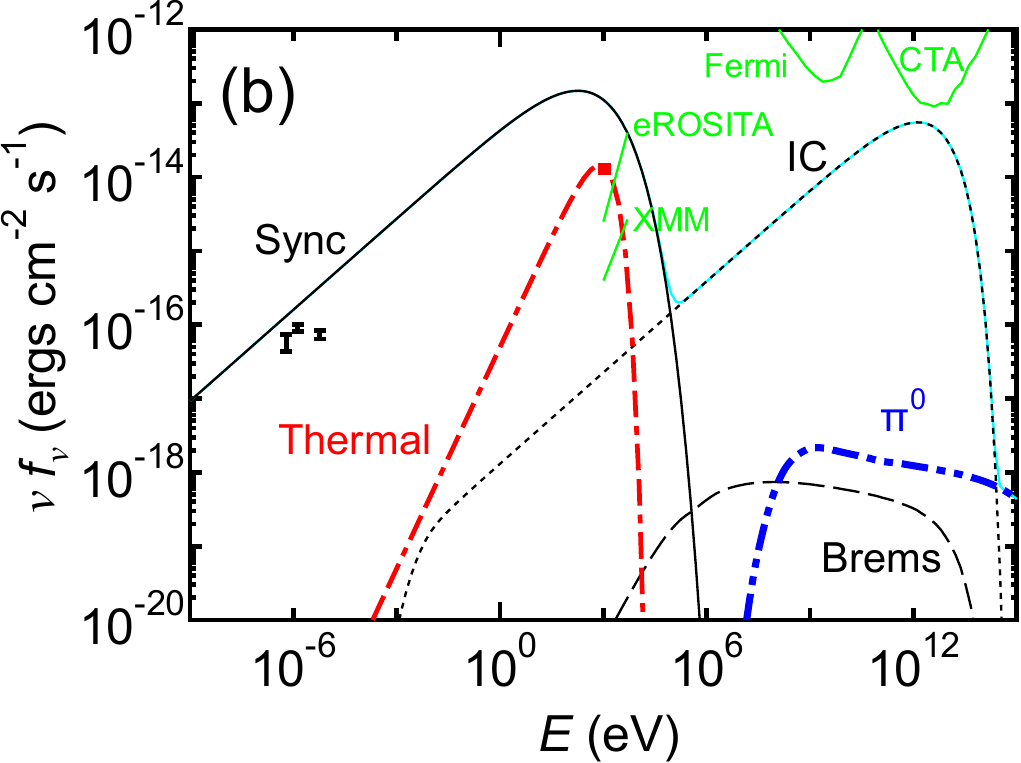} 
 \end{center}
\caption{Same as figure~\ref{fig:ORC4} but for ORC4 without radiative cooling. (a) Current time ($t_0=130$~Myr). (b) $t=0.2\: t_0=25$~Myr.}.\label{fig:ORC4n}
\end{figure*}
\begin{figure*}[ht!]
 \begin{center}
  \includegraphics[width=8cm]{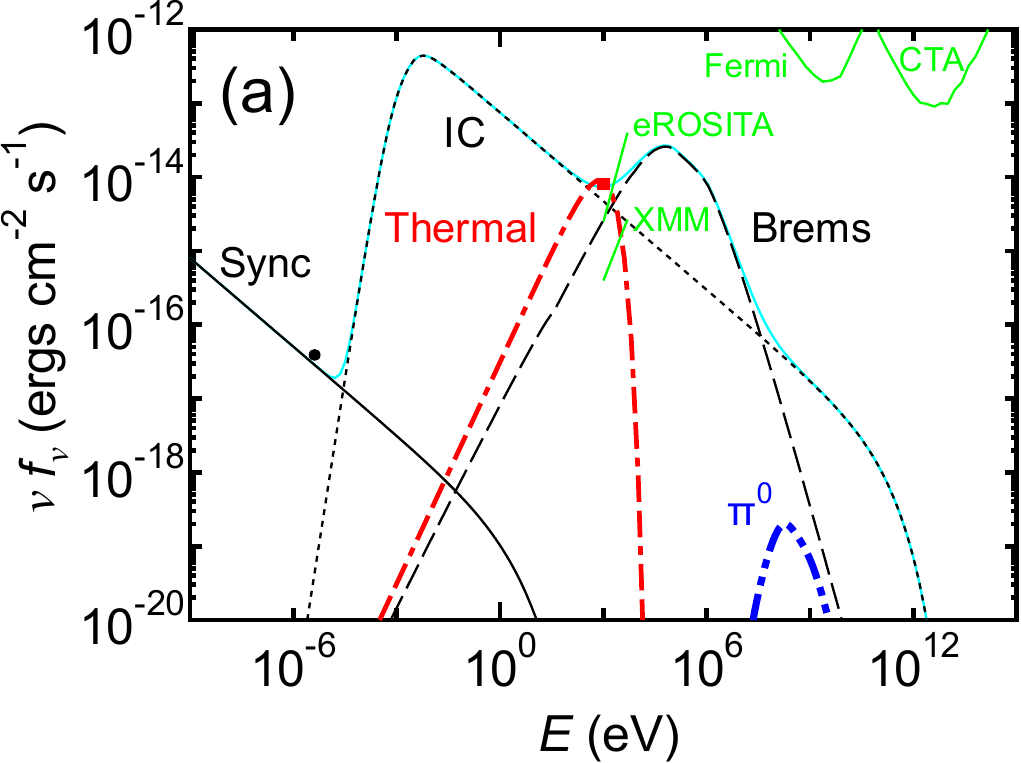} 
  \includegraphics[width=8cm]{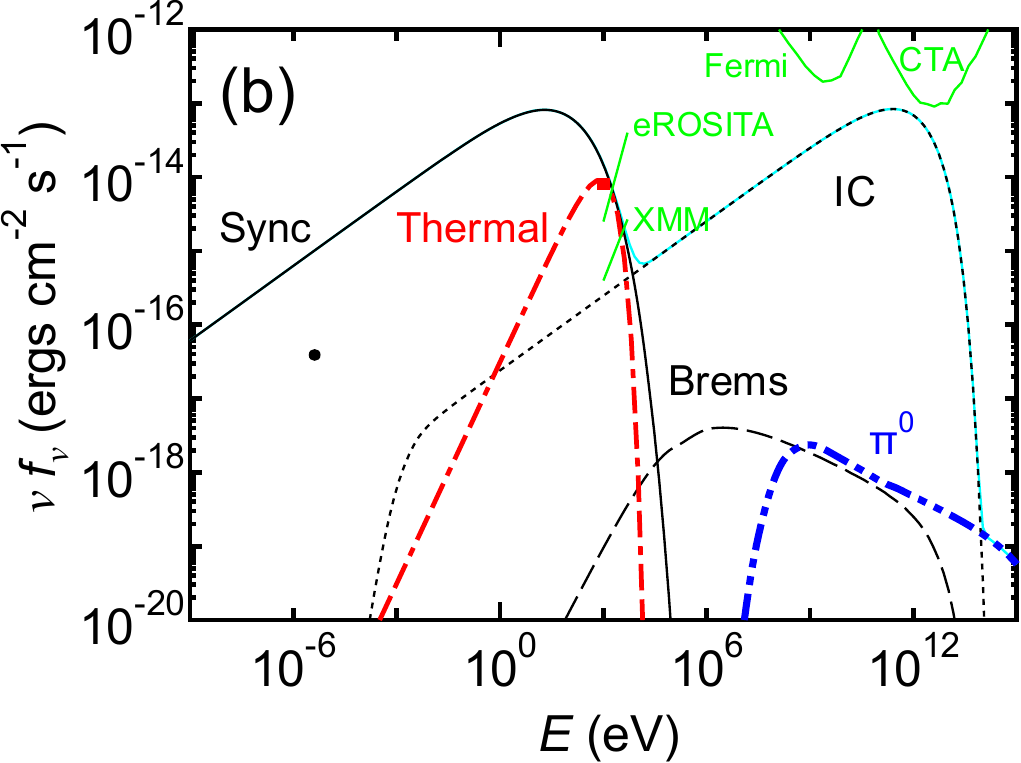} 
 \end{center}
\caption{Same as figure~\ref{fig:ORC4} but for ORC1 without radiative cooling. (a) Current time ($t_0=220$~Myr). (b) $t=0.2\: t_0=44$~Myr.}.\label{fig:ORC1n}
\end{figure*}

\section{Discussion}

\subsection{Implications of the model}

Our OGRE model indicates that the ejected energy is ${\cal E}_0\sim 10^{60}$~erg and the shock age is $\sim 10^8$~yr for both ORC4 and 1.
Assuming that the energy was supplied by supernovae on a timescale of $\sim 10^7$~yr, $\sim 100$ supernova explosions per year must have occurred $\sim 10^8$~yr ago, which seems extreme considering the star formation history of the central galaxies \citep{2022MNRAS.513.1300N,2023arXiv231015162C}. 
\citet{2023arXiv231015162C} suggested that ORC4 could have been produced by a starburst-driven outflow about 1~Gyr ago (see also \cite{2023arXiv231206387R}). However, with the much longer age, the shock velocity $V_s$ in this case must be much lower than that in our model, leading to a smaller ${\cal E}_0$ that may be inconsistent with the observed synchrotron luminosity.
Thus, AGNs at the galactic center may be a more promising candidate for the power source. The ionized gas and its kinematics around the central galaxy of ORC4 found by \citet{2023arXiv231015162C} could have been excited by AGN activity about 0.1~Gyr ago.
Such AGN outbursts may be responsible for widespread metals in intergalactic space \citep{2008PASJ...60S.343F}.

Figures~\ref{fig:ORC4} and~\ref{fig:ORC1} show that the SEDs of the younger OGREs are not very different from those of older ones corresponding to the currently observed ORCs. In the radio band ($\sim 10^{-5}$~eV or $\sim$~GHz), the differences in flux is less than a factor of 10. Assuming that the number of OGREs is proportional to their age, the probability of finding younger OGREs is expected to be lower
\footnote{Since the evolution of an OGRE follows a power law (equation~(\ref{eq:Rs})), the evolution rate decreases gradually (e.g. $dR_s/dt$). However, it is constant in logarithmic space ($d\log R_s/d\log t$). Thus, it is natural to assume that the OGREs with logarithmic radius from $\log R_s(t)$ to $\log R_s(t) + a$ are practically of the same age, where $a=d(\log R_s)=dR_s/R_s$ is a small dimensionless constant. Here we assume that $N_{\rm O}(t)dt$ is the number of OGREs aged between $t$ and $t+dt$. If $dt/t\ll 1$, then the OGREs aged between $t$ and $t+dt$ can be considered to be of the same age, since $dt/t\propto dR_s/R_s$. If OGREs are born at a constant rate, then $N_{\rm O}(t)$ must be constant. Assuming $dt/t=b$, where $b$ is a small constant, the number of OGREs at age $t$ is $N_{\rm O}(t)dt \propto N_{\rm O}(t)bt \propto t$.}.
Their smaller size could also make their detection more difficult.

In figures~\ref{fig:ORC4} and~\ref{fig:ORC1}, for comparison, we show the thermal emission from the hot gas of NGC~4636 placed at the redshifts of the ORCs; the flux is estimated from the observed value at 0.2--2.4~keV at its distance of 14.7~Mpc \citep{2004A&A...423..449P,2021MNRAS.506.2030M}. As the synchrotron and IC emission are both fainter than the thermal emission, detection of ORCs in X-rays may be difficult. The IC emission peaks in the MeV bands, but their detection with near-future instruments may also be difficult.

\subsection{Ineffective radiative cooling}

In the context of a model for ORCs as synchrotron emission from electrons accelerated at virial shocks,
\citet{2023arXiv230917451Y} 
suggested that the diffusion coefficient of electrons downstream of the shock may need to be relatively large, in order to explain the observed width of ORCs.
If the relevant diffusion coefficient can be similarly large in our model, the electrons may cool adiabatically before cooling radiatively, since our model is a spherical explosion model and the electrons scattered toward the inner region are affected by gas expansion. Thus, we also consider cases where radiative cooling is very inefficient and the energy distribution of electrons is always given by equation~(\ref{eq:Ni(E)}).

We show the results for ORC4 in figure~\ref{fig:ORC4n}. We assume that $\alpha_{\rm obs} = \alpha=0.92$, implying $\mathcal{M}=2.4$ (equation~(\ref{eq:alpha})) and $x=2\alpha + 1 = 2.84$. The current shock radius is $R_s=200$~kpc \citep{2022MNRAS.513.1300N} and the velocity is $V_s=876\rm\: km\: s^{-1}$. Since the value of $V_s$ is much smaller than that in section~\ref{sec:ORC4}, higher CR acceleration and emission efficiencies are required. Thus, we set $B_2=4.5\:\rm \mu G$ or our maximum value (table~\ref{tab:par}).
We take $\epsilon=0.1$ and  $K_{\rm ep}=0.05$ to be consistent with the observed radio luminosity.
Figure~\ref{fig:ORC4n}(a) shows the current SED for ORC4. The explosion energy is ${\cal E}_0=8.1\times 10^{59}$~erg from equation~(\ref{eq:E0}). The time since the explosion is estimated to be $t_0=130$~Myr from equation~(\ref{eq:Rs}). The maximum proton and electron energies are $E_{{\rm max},p}=48$~PeV and $E_{{\rm max},e}=9.2$~TeV, respectively (equations~(\ref{eq:Emaxp}) and~(\ref{eq:Emaxe})).
Figure~\ref{fig:ORC4n}(b) shows the SED at $t=0.2\: t_0=25$~Myr. The radius is $R_s=81$~kpc (equation~(\ref{eq:Rs})) and the velocity is $V_s=1780\rm\: km\: s^{-1}$. The maximum energies are $E_{{\rm max},p}=127$~PeV and $E_{{\rm max},e}=21$~TeV. The synchrotron and IC spectra are much harder ($\alpha=0.59$ and $x=2.18$) than those in figure~\ref{fig:ORC4n}(a) due to the higher Mach number of the younger shock ($\mathcal{M}=4.9$), which can be estimated from equation~(\ref{eq:alpha}). The synchrotron spectrum peaks in the soft X-ray band and the IC spectrum peaks in the TeV band.  

The results for ORC1 are presented in figure~\ref{fig:ORC1n}. We assume $\alpha_{\rm obs}=\alpha=1.4$, which means $\mathcal{M}=1.8$ (equation~(\ref{eq:alpha})) and $x=3.8$. The current shock radius is $R_s=260$~kpc \citep{2022MNRAS.513.1300N} and the velocity is $V_s=655\rm\: km\: s^{-1}$. 
We assume $B_2=4.5\rm\:\mu G$ as ORC4 (table~\ref{tab:par}). We set $\epsilon=0.3$ and $K_{\rm ep}=0.1$ so that the predicted radio flux matches that observed. The explosion energy is ${\cal E}_0=6.8\times 10^{59}$~erg (equation~(\ref{eq:E0})) and the time since the energy ejection is estimated to be $t_0=220$~Myr (equation~(\ref{eq:Rs})).
The maximum energies are $E_{{\rm max},p}=29$~PeV and $E_{{\rm max},e}=5$~TeV  (equations~(\ref{eq:Emaxp}) and~(\ref{eq:Emaxe})).
Figure~\ref{fig:ORC1n}(a) shows the current SED. The IC spectrum peaks in the infrared band and the non-thermal bremsstrahlung spectrum peaks in the MeV band. 
Figure~\ref{fig:ORC1n}(b) shows the SED at $t=0.2\: t_0=44$~Myr. The radius is $R_s=106$~kpc (equation~(\ref{eq:Rs})) and the velocity is $V_s=1330\rm\: km\: s^{-1}$. The maximum energies are $E_{{\rm max},p}=89$~PeV and $E_{{\rm max},e}=15$~TeV. The SED is similar to that of ORC1 in the past (figure~\ref{fig:ORC4n}(b)), due to the larger Mach number ($\mathcal{M}=3.6$) and the harder spectra ($\alpha=0.66$ and $x=2.33$),

Since the synchrotron emission of the young OGREs in the X-ray band exceeds the thermal emission (figures~\ref{fig:ORC4n}(b) and~\ref{fig:ORC1n}(b)), the former may be detectable and distinguishable from the latter through the differences in morphology (shell-like versus centrally concentrated).
Note that the thermal flux can be much smaller than shown in figures~\ref{fig:ORC4n} and~\ref{fig:ORC1n} if the hot gas in the inner regions of the central galaxy has been expelled by the explosion.
The diameter of the young OGREs is $\sim 40''$ at $z\sim 0.5$, which is larger than the angular resolution of eROSITA ($\sim 15''$; \cite{2021AandA...647A...1P}). Thus, they may be observable as “odd X-ray circles”.

\subsection{Caveats}

Our model includes parameters such as $kT$, $\omega$, $B_2$, $\epsilon$, and $K_{\rm ep}$ that are difficult to identify uniquely from the currently available observational information on ORCs. The numbers we have chosen here may be only one set out of various possible ones. Future multi-band observations could be a powerful means to constrain such quantities. If the prior probability distributions of these parameters can be reasonably estimated, Markov chain Monte Carlo (MCMC) methods could also be effective.

Because our model is rather simple, some observed features of ORCs cannot be reproduced. For example, ORC1 has a multiple ring structure \citep{2022MNRAS.513.1300N}. This might be explained if the shock is produced by outflows with a more complex, bipolar structure instead of the spherically symmetry assumed here.

Given that the observational information on ORCs is still very limited (radio power, size, spectral index), a crucial step in our discussion was the application of equation~(\ref{eq:alpha}) to infer the shock Mach number from the observed radio spectral index.
Our current understanding of DSA is still incomplete, and such simple relations between $\alpha$ and $\mathcal{M}$ may not hold for shocks with sufficiently high Mach numbers due to nonlinear effects \citep{2001RPPh...64..429M} or relativistic shocks \citep{2015SSRv..191..519S}.
For low Mach number shocks in clusters, observations have reported discrepancies between radio and X-ray derived Mach numbers \citep{2013PASJ...65...16A,2015A&A...582A..87A,2021MNRAS.506..396W}, which may be due to projection effects \citep{2013ApJ...765...21S,2015ApJ...812...49H}. Thus, there is uncertainty in the assumed relation between the Mach number $\mathcal{M}$ and the spectral index $\alpha$ (equation~(\ref{eq:alpha})). If the radio index $\alpha$ does not correctly reflect the Mach number $\mathcal{M}$, the current SED may be different from our predictions (figures~\ref{fig:ORC4}(a) and \ref{fig:ORC1}(a), and figures~\ref{fig:ORC4n}(a) and \ref{fig:ORC1n}(a)).
However, we predict that young OGREs should always have similar SEDs (figures~\ref{fig:ORC4}(b) and \ref{fig:ORC1}(b), and figures~\ref{fig:ORC4n}(b) and \ref{fig:ORC1n}(b)). Moreover, recent theoretical studies suggest that the spectral index obtained from radio observations is relatively immune to projection effects \citep{2021MNRAS.506..396W}.

\section{Conclusions}

In the present study, we propose that ORCs are the remnants of explosive galactic outflows, which we call OGREs. A significant energy release is assumed to have occurred in the past in the central region of a galaxy, producing an outward shock that accelerates cosmic rays. Assuming reasonable parameters for the surrounding medium, the observed properties of ORCs, namely their spectral index, dimensions, and energy output, require an energy input on the order of about $10^{60}$ erg, implying AGNs as plausible progenitors.

We have calculated the SEDs associated with the OGREs, along with their temporal evolution, including contributions from synchrotron, IC, and bremsstrahlung emission from electrons, and pion decay emission from protons. We found that the SEDs of young OGREs are similar to older ones corresponding to the currently observed ORCs when radiative cooling is effective. Given the expected rarity and smaller dimensions of young OGREs, their detection may be more difficult than that of older and larger ones seen as ORCs. Conversely, in scenarios where radiative cooling is less effective, young OGREs may manifest observable X-ray signatures.

\begin{ack}
We thank the reviewer for helpful comments that improved the paper. We also thank H. Akamatsu for useful comments.
This work was supported by JSPS KAKENHI Grant Number JP22H00158, JP22H01268, JP22K03624, JP23H04899 (Y.F.), JP22K03686 (N.K).
\end{ack}

\appendix 
\section*{Energy distribution of CRs}

Assuming that the energy distribution of protons is given by
\begin{equation}
 N_p(E) = A_p E^{-x}e^{-E/E_{{\rm max},p}}\:,
\end{equation}
where $A_p$ is the normalization (see equation~(\ref{eq:Ni(E)})). Since the total energy of the proton is $\epsilon{\cal E}_0$, the normalization is
\begin{equation}
 A_p = \epsilon{\cal E}_0/\int_{E_{{\rm min},p}}^{\infty}E^{-x}e^{-E/E_{{\rm max},p}}
\end{equation}

If radiative cooling is ignored, the energy distribution of the electrons is given by
\begin{equation}
 N_e(E) = A_e E^{-x}e^{-E/E_{{\rm max},e}}\:,
\end{equation}
where $A_e$ is the normalization, given by
\begin{equation}
 A_e = K_{\rm ep}A_p
\end{equation}

%

\end{document}